\documentstyle[preprint,aps]{revtex}

\begin{document}
\draft
\preprint{ UOM-NPh-hqp/1-99 (revised)}
\title{Colour-straight four-quark operators and lifetimes of
beautiful hadrons}
\author{S. Arunagiri}        
\address{Department of Nuclear Physics, University of
Madras,\\
Guindy Campus, Chennai 600 025, Tamil Nadu, INDIA}
\date{\today}
\maketitle
\begin{abstract}
Using the relation between the harmonic oscillator wave
function and the light quark scattering form factor, the
expectation values of colour-straight four-quark operators
are evaluated and found to be directly proportional to the
cubic power of the oscillator strength. It is predicted that
the ratio ${\tau}$(${\Lambda_b}$)/${\tau}$(B) ${\approx}$
0.79(0.84) due to the factorizable (nonfactorisable) piece,
against the experimental 0.79 ${\pm}$ 0.06. Notwithstanding
the numerical prediction, the present study shows that the
four-quark operators play a role as far the lifetimes of
b-flavoured hadrons are concerned.
\end{abstract}

\section{Introduction}
In the description of heavy hadron decays by heavy quark
expansion (HQE), the preasymptotic effects appearing at
next-to-leading order and beyond are vital in predicting
the decay properties accurately. These effects are due to
the operators of dimensions D = 5 and 6. At 1/${m_Q^2}$,
the operators are suppressing the leading order. The
evaluation of D = 5 operators which describe the motion
of the heavy quark inside the hadron and the chromomagnetic
interaction is definite. The estimation of the D = 6 operators
which are four-quark operators (FQO) containing both heavy and light
fields is based on the vacuum insertion assumption for mesons
and on the quark models for baryons. Though their effects are
negligible as the heavy quark's volume occupation is of the
order of ${(\Lambda_{QCD}/m_Q)^3}$ but are considerably
enhanced due to partial compensation by the four-quark
phase-space, these operators are predicting the lifetime
differences in the world of the hadrons
of given flavour quantum number. Therefore, accurate value
of the FQO is necessary due to the confrontation
existing between theory and experiment over the hadronic
properties like the experimentally smaller than theoretically
expected lifetime of ${\Lambda_{b}}$ and the theoretically
smaller than experimentally predicted semileptonic branching
ratio of B-meson. Theoretically upto order (1/m$_b$$^2$)
\cite{bigi95,neubert98}
\begin{eqnarray}
{\tau(B^+) \over {\tau(B^0)}} &=& 1+0.05~O
\left ({f_B \over {200 MeV}}\right )^2 \nonumber\\
{\tau(B_s) \over {\tau(B^0)}} &=& 1+0.01\nonumber\\
{\tau(\Lambda_b) \over {\tau(B^0)}} &=& 0.9
\end{eqnarray}
whereas experimentally \cite{caso98},
\begin{eqnarray}
{\tau(B^+) \over {\tau(B^0)}} &=& 1.04 \pm 0.04\nonumber\\
{\tau(B_s) \over {\tau(B^0)}} &=& 0.99 \pm 0.05\nonumber\\
{\tau(\Lambda_b) \over {\tau(B^0)}} &=& 0.79 \pm 0.06
\end{eqnarray}

The agreement among the B mesons is as expected of but not
so between the ${\Lambda_b}$ and B. The later issue continues
to be central point of the physics of heavy quark hadrons.
It is suspected that the explanation for these discrepancies
within the HQET framework is hidden in the
yet-not-satisfactorily-understood FQO.

As regards the evaluation of the FQO,
there were two works \cite{neubert97,rosner96} which attempted
to explain substantially the enhanced decay rate of
${\Lambda_b}$, whereas the work of P. Colangelo and F. De Fazio
\cite{colan96} which is QCD sum rules based prediction leads
to conclude that the reason for the smaller lifetime is not
due to FQO. In Ref.\ \onlinecite{neubert97},
the authors evaluated the FQO parameterising
the matrix elements in terms of hadronic parameters which are
not practically known. But the parameters have been calculated
using QCD sum rules \cite{yang98}. However, the prediction
is not able to account for the lifetime difference between
${\Lambda_b}$ baryon and B meson. On the other hand, the author
of Ref.\ \onlinecite{rosner96} used quark model and accounted
for the FQO for 13${\%}$ of the required
enhancement in the ${\Lambda_{b}}$ decay rate. The above estimation
used yet to be confirmed result of DELPHI collaboration
\cite{abreu95} on the mass splitting of ${\Sigma_b^*}$ and
${\Sigma_b}$. The same method has been modified by taking the
logarithmic dependence of the wave function at the origin and
this explains the difference between the decay rates of
B meson and ${\Lambda_{b}}$ baryon to the extent of 40${\%}$ 
\cite{arunagiri98}. Since the striking point in the evaluation
of the FQO is not yet obtained to clear the
situation in one way or the other, it is important and
interesting to explore other avenues to estimate the
four-quark matrix elements.

In this paper, we adopt the colour-straight formalism approach
of \cite{pirjol98} to evaluate the expectation values of
the four-quark matrix elements. On the specific choice of the
harmonic oscillator wavefunction model for the form factor and
slightly different potentialr for meson and baryon,
it is found that
\begin{eqnarray}
{\tau(B^+) \over {\tau(B^0)}} &=& 1.00 (1.03) \nonumber\\
{\tau(B_s) \over {\tau(B^0)}} &=& 1.00 (1.02) \nonumber\\
{\tau(\Lambda_b) \over {\tau(B^0)}} &=& 0.79 (0.84)
\end{eqnarray}
where the values given within brackets are due to non-factorisable
part of the FQO. These values are in agreement with the data, eq. (2).

In a recent work \cite{pirjol98}, Pirjol and Uraltsev discussed
the four-fermion operators on certain quantum mechanical basis.
In the nonrelativistic quark theory, the wave function density
and diquark density are related to the associated operator 
\begin{equation}
(\bar b_i \Gamma_b b^i)(\bar q_j \Gamma_q q^j),
\end{equation}
where the ${\Gamma_{b, q}}$ are arbitrary Dirac structures,
through
\begin{equation}
{1 \over {2M_B}} <B| (\bar b b)(\bar q q)| B> = |\Psi (0)|^2
\end{equation}
\begin{equation}
{1 \over {2M_{\Lambda_b}}} <{\Lambda_b}| (\bar b b)(\bar q q)
|{\Lambda_b}>
= \int d^3y |\Psi (0, y)|^2
\end{equation}
for meson and baryon respectively. The operators in Eq. (4)
are colour singlet. As colour flows freely for these operators,
they are called colour-straight. The expectation values of
the operators in Eq. (4) are related to the observable, the
transition amplitude of the light quark scattering off the
heavy quark. Thus the determination is based on the
knowledge of the light quark scattering form factor. 

The wave function at the origin, in momentum representation,
is given by
\begin{equation}
\Psi(0) = \int {d^3{\bf p} \over {(2 \pi^3)^3}} \Psi({\bf p}).
\end{equation}
The transition amplitude is then the Fourier transform of the
light quark density distribution:
\begin{equation}
F({\bf q}) = {1 \over {2M_H}}<H_b({\bf q})|\bar q q(0) |H_b(0)> = 
\int d^3{\bf x} \Psi^*({\bf x}) \Psi({\bf x}) e^{i{\bf q}{\bf x}}.
\end{equation}
Integrating over all {\bf q} yields the expectation value:
\begin{equation}
\int {d^3{\bf q} \over {(2 \pi)^3}} F({\bf q}) = |\Psi(0)|^2 = 
{1 \over {2M_H}}<H_b|\bar b b \bar q q(0) |H_b> 
\end{equation}
For any Dirac structure ${\Gamma}$, the light quark current
density and the light quark transition amplitude are given by:
\begin{equation}
J_\Gamma({\bf x}) = \bar q \Gamma q({\bf x}); ~~~A_\Gamma({\bf q}) = 
{1 \over {2M_H}}<H_b({\bf q})|J_\Gamma(0) |H_b(0)> 
\end{equation}
where the J${_\Gamma}$(0) is gauge invarinat operator and not
required to be a bilinear. Thus, for spin-singlet operators,
we have
\begin{equation}
<H({\bf q})|\bar b b  J_\Gamma (0) |H_b(0)> = 
\int {d^3{\bf q} \over {(2 \pi)^3}} <H_b({\bf q})|J_\Gamma(0) |H_b(0)> 
\end{equation}
And, for spin-triplet operators, similarly
\begin{equation}
<\ H({\bf q})|\bar b {\bf \sigma_k} b \bar J_\Gamma (0) |H_b(0)> = 
\int {d^3{\bf q} \over {(2 \pi)^3}} <S_k \ H_b({\bf q})
|J_\Gamma(0) |H_b(0)>
\end{equation}
with S/2 being the b-quark spin operator and 
\begin{equation}
|S_k \ H_b> = \int d^3{\bf x} \bar b \sigma_k b({\bf x}) |\ H_b>
\end{equation}
Equations (11) and (12) represent the general structure of four-quark 
operators. The above operators are local  as required for by
the HQE in the sense that the light quark operators enter
at the same point as the heavy b-quark operators. These relations,
Eqs. (11) and (12), hold equally for different initial and final
hadrons having different momenta smaller than m${_b}$.

Equation (4)  resolves into spin-singlet and spin-triplet operators
for ${\Gamma_b}$ = 1 and ${\Gamma_b}$ = ${{\bf \gamma} \gamma_5
(= {\bf\sigma})}$ respectively. The light quark elastic scattering
is described by the form factor F(q$^{2}$). In Ref.
\onlinecite{pirjol98}, the exponential ansatz and the two pole
anstaz are used for the form factor. Both of them lead to a
determination which make no difference for meson and  baryon.
This cannot truly be the case to represent ${|\Psi(0)|^2}$,
the measure of the expectation values of FQO,
for a baryon and a meson.

In the next section, the choice of the representation, harmonic
oscillator wave function for the form factor and the potentials
are discussed. The evaluation of the expectation values of the
factorisable part of the FQO is presented in Sec. III. Corresponding
non0factorisable part is given in Sec. IV. Estimation of the lifetimes
ratio and the conclusion are given in Sec. V and VI respectively.

\section{HARMONIC OSCILLATOR WAVE FUNCTION MODEL FOR FORM FACTOR}
\label{sec:level1}
As will be discussed in the following sections, the expectation
values of the colour-straight operators are parameterised in
terms of a single form factor  for both B-meson and ${\Lambda_b}$
baryon. The extraction of the form factor involves assumption of
a function such that it satisfies the constraints on the form factor
that F(q${^2}$ = 0) is equal to the corresponding charge of
the hadron. Then the form factor has to be extrapolated into
the region where q${^2}$ ${>}$ 0. We take the hadronic wave
function of ISGW harmonic oscillator model \cite{grin89} for
the form factor.

The wave functions of ISGW model are the eigenfunctions of
orbital angular momentum L = 0 satisfying the overlapping
integral
\begin{equation}
I({\bf q}) = \int r^2 dr \Psi_f^* (r) \Psi_i (r) j_0 ({\bf q}r). 
\end{equation}
The overlapping integral can be equated to the form factor.
Hence for diffrent initial and final hadrons
\begin{equation}
F({\bf q}^2) = N^2 exp[-q^2/2(\beta_f^2+\beta_i^2)] 
\end{equation}
where N is the normalisation constant given by 
${[2 \beta_f \beta_i/(\beta_f^2+\beta_i^2)]^{3/2}}$ and
${\beta}$'s are oscillator strengths. For same initial
and final hadrons, the transition amplitude is
\begin{equation}
\int {d^3{\bf q} \over {(2 \pi)^3}} F(q^2) =
{\beta^3 \over {4\pi^{3/2}}} =
|\Psi(0)|^2
\end{equation}
From the above equation, which is the central point of dicussion
of this paper, it is obvious that the transition amplitude and
hence the expectation values of four fermion operators are
proportional to the third power of the oscillator strength
of the hadron.

The calculation of ${\beta}$'s can be made using the QCD inspired 
potential. In the present calculation, we use the potential for
B-meson containing the Coulomb, confining and a constant terms: 
\begin{equation}
V(r) = {a \over {r}} + br + c
\end{equation}
For a = -0.508, b = 0.182 GeV${^2}$ and c = -0.764 GeV, and
for quark masses m${_q}$ = 0.3 GeV (treating m${_u}$ =
m${_d}$), m$_s$ = 0.5 GeV and m${_b}$ = 4.8 GeV, using
variational procedure with the wave function given
in position space as, 
\begin{equation}
\Psi(0)_{1s} = {\beta^{3/2} \over \pi^{3/4}} e^{\beta^2r^2/2}
\end{equation}
${\beta}_{B_q}$ = 0.4 GeV and ${\beta}_{B_s}$ = 0.44 GeV. 

For ${\Lambda_b}$ baryon, following the similar procedure
but for the potential of the form
\begin{equation}
V(r) = {1 \over {2}} \left ({a \over {r}} + br +
\beta r^2 + c \right )
\end{equation}
where the r${^2}$ term is a harmonic oscillator term justifying
the consideration that ${\Lambda_b}$ be a two body system and
the same wave function of Eq. (18), one gets ${\beta_{\Lambda_b}}$
= 0.72 GeV and for ${\beta_{\Xi_b}}$ = 0.76 GeV for the values
of mass of the diquark system 0.6 and 0.8 GeV respectively.
The large value for the $\beta_{|Lambda}$ is due to the
presence of the O(r$^2$) term in the potential. Otherwise,
the value of $\beta_{\Lambda}$ is no different than that of B$_s$.
These values are used in the subsequent calculations in this
paper. A comment is in order on the choice of the
same wave function for baryon as for meson: In the usual procedure, 
the ground state wave function for baryon is
\begin{equation}
\Psi_{ground} \approx e^{-\alpha^2(r_\lambda^2+r_\rho^2)/2}
\end{equation}
where r$_{\lambda, \rho}$ are the internal coordinates for three
body system. Due to the idea of considering the ${\Lambda_b}$ as
a system containing the bound state of light quarks and a heavy
quark, the separation of the two light quarks which make the
bound state is treated negligibly. This allows then that the
baryon is a system of two body. It is a reasonable
approximation only. The difference between a meson and a baryon
is essentially due to the value of the oscillator strength. 

\section{EXPECTATION VALUES OF THE colour-straight OPERATORS}
We evaluate the expectation values of the colour-straight operators
only for the vector and axial-vector currents. Nevertheless the other 
currents can also be studied in the same fashion. Both the currents
are possible for B-meson while axial currents vanish for
${\Lambda_b}$ baryon due to the light degrees of freedom which
constitute a spinless bound state.

Essentially there is no difference between the exponential
ansatz and the harmonic oscillator wave function in representing
the behaviour of the form factor but they differ while fixing
the scale: in the former case, the hadronic scale of one GeV
is used whereas in the latter the same has been fixed by
solving the Schrodinger equation. The two pole anstaz is based
on the well founded experimental values. Basically the use of the
harmonic oscillator wave function of the constiuent quark model
is an alternate picture but in the very same lines of the two
ansatz.

Hereinafter the operators are referred to by the following
notaion: for meson
\begin{equation}
<O_{V, A}^q>_H = <H|\bar b \Gamma_{V, A} b \bar q \Gamma_{V, A} q
|H>;~~~~~
<T_{V, A}^q>_H = <H|\bar b \Gamma_{V, A} t^a b \bar q \Gamma_{V, A} t^a q
|H>
\end{equation}        
where $\Gamma_{V, A} = \gamma_{\mu}, \gamma_{\mu} \gamma_5 $ and 
${t^at^b = 1/2-1/2N_c}$. 
In what follows, q stands for u and d quarks and s for strange
quark. And ${<O_{V, A}>, <T_{V, A}>}$ will be respectively
denoted as ${\omega_{V, A}, \tau_{V, A}}$. For baryon, $<O(T)_V>$
correspond to $\lambda_{1, 2}$ respectively.

\subsection{B-meson} 
The parametrisation of the matrix element of the colour-straight
operators for vector current is
\begin{equation}
<B(q) | J_{\Gamma_V} |B(0)> = -v_{\mu} F_B(q^2)
\end{equation}
with the constraints F${_B}$(0) = 1 for valence quark current
and F${_B}$(0) = 0 for sea quark current. The former is relevant
for the b-meson composition of quarks b${\bar q}$. Then the
corresponding transition amplitude is
\begin{equation}
A_{\Gamma_V}^B = <B(q) | O_V^q  |B(0)> =
-v_\mu \int {d^3 q \over {(2 \pi)^3}}F_B(q^2)
\end{equation}
Under isospin SU(2) symmetry,
\begin{equation}
<B^- | O_V^q |B^-> = - 2.88 \times 10^{-3}~GeV^3;~~~~~~
<B^-|T_V^q|B^-> = -9.61 \times 10^{-4} GeV^3
\end{equation}
For B${_s}$, we have 
\begin{equation}
<B_s | O_V^q |B_s> = - 3.83 \times 10^{-3}~GeV^3;~~~~~
<B_s | T_V^q |B_s> = - 1.28 \times 10^{-3}~GeV^3
\end{equation}
If the case of violation of isospin symmetry, and SU(3)
flavour symmetry, is considered, then there comes Cabibbo
mixing of eigenstates for d and s quarks among themselves.
That is, $d^{\prime} = d cos\theta_c+s sin\theta_c$
and $s^{\prime} = s cos\theta_c+d sin\theta_c$. This we do not
consider here. 

For axial current there are two form factors which are related 
to one another due to conservation of the axial current,
${{\partial_\mu} J_{\mu 5}}$ = 0, in the chiral limit.
By the Goldberger-Treiman relation
\cite{don92} which equates axial charge form factor to the coupling 
g${_{B^*B \pi}}$ at q${^2}$ = 0, the operators involving
axial-currents are estimated in terms a single form factor.
Thus, given the value of the coupling g, the extraction of the
transition amplitude is similar to the B-meson case.

Making use of 
\begin{equation}
(S_b \epsilon)|B(q)> = |B^*(q,\epsilon)>
\end{equation}
and Eq.(12), the expectation values for the axial vector
currents are given by
\begin{equation}
<B(q, \epsilon)^* | \sum_{q = u,d,s}
\bar q \gamma_\mu \gamma_5 q |B(0)> =
\epsilon_\mu^* G_1^{(0)}(q^2)-(\epsilon^*q)q_\mu
G_0^{(0)}(q^2)
\end{equation}
\begin{equation}
<B(q, \epsilon)^* | \bar q \lambda^a  \gamma_\mu \gamma_5 q |B(0)> = 
\lbrace\epsilon_\mu^* G_1^{(0)}(q^2)-(\epsilon_\mu^*q)q_\mu G_0^{(0)}
(q^2)\rbrace\lambda_{ij}^q 
\end{equation}
Finally the following equality leads the absence of the structure 
(${\epsilon^*q)}v{_\mu}$
\begin{equation}
<B^*(q,\epsilon)|j_{\mu 5}(0)|B(0)>^* =
<B(0)|j_{\mu 5}(0)|B^*(q,\epsilon)>
= <B^*(0,\epsilon)|j_{\mu 5}(0)|B(q)>
\end{equation}
Following the Goldberger-Treiman relation, we have
\begin{equation}
G_1(q^2) = q^2 G_0(q^2) = g e^{-q^2/{4\beta^2}}
\end{equation}
Correspondingly, the expectation values ars
\begin{equation}
<B^-|O_A^q|B^-> =  8.63\times 10^{-5}~GeV^3;~~~~~
<B^-|T_A^q|B^-> =  2.88\times 10^{-5}~GeV^3
\end{equation}
\begin{equation}
<B_s|O_A^s|B_s> =  1.15 \times 10^{-4}~GeV^3;~~~~~
<B_s|T_A^s|B_s> =  3.84 \times 10^{-5}~GeV^3
\end{equation}
We have taken in the above esitmates the value g = - 0.03
\cite{bel95,dos96}.

\subsection{${\Lambda_b}$ baryon}

For ${\Lambda_b}$ baryon, treating u and d quarks equally, 
\begin{equation}
<\Lambda_b|O_V^q |\Lambda_b> = -1.69\times 10^{-2}~GeV^3;~~~~~
<\Lambda_b|T_V^q |\Lambda_b> = -5.64\times 10^{-3}~GeV^3
\end{equation}
In the case of ${\Xi_b}$, we have, 
\begin{equation}
<\Xi_b|\sum_{q^\prime = u,d,s} O_V^{q^\prime} |\Xi_b > = -2.01 
\times 10^{-2}~GeV^3;~~~~~
<\Xi_b|\sum_{q^\prime = u,d,s} O_V^{q^\prime} |\Xi_b > = -6.72 
\times 10^{-3}~GeV^3
\end{equation}
There are corrections additionally to form factors due to
charge radius. The same can be ignored as we are looking
at the wave function density at the origin.

\section{Non-factorisable part of the FQO}
The nonfactorisable part of the FQO come in four.
The following is one of the ways of parameterising them
\cite{neubert97}.
\begin{equation}
{1 \over {2M_B}}<B|(\bar b q)_{V-A} (\bar q b)_{V-A}|B> = 
{\bar f_B^2} M_B B_1/2
\end{equation}
\begin{equation}
{1 \over {2M_B}}<B|(\bar bt^aq)_{V-A} (\bar qt^ab)_{V-A}|B> = 
{\bar f_B^2} M_B \epsilon_1/2
\end{equation}
\begin{equation}
{1 \over {2M_B}}<B|(\bar b q)_{S-P} (\bar q b)_{S-P}|B> = 
{\bar f_B^2} M_B B_2/2
\end{equation}
\begin{equation}
{1 \over {2M_B}}<B|(\bar bt^aq)_{S-P} (\bar qt^ab)_{S-P}|B> = 
{\bar f_B^2} M_B B_2/2
\end{equation}
where $B_{1, 2}$ and $\epsilon_{1, 2}$ are hadronic parameters.
They are related to $w_{V, A}$ and $\tau_{V, A}$ which are
the expectation values of the operators
$O_{V, A}$ and $T_{V, A}$ as defined earlier.
\begin{equation}
{\bar f_B^2} M_B B_1 = \phi_1 = 4(\tau_V+\tau_A)+{2 \over {N_C}}
(\omega_V+\omega_A)
\end{equation}
\begin{equation}
{\bar f_B^2} M_B B_2 = \phi_2 = -2(\tau_V-\tau_A)-{1 \over {N_C}}
(\omega_V-\omega_A)
\end{equation}
\begin{equation}
{\bar f_B^2} M_B \epsilon_1 = \rho_1 = -{2 \over {N_C}}
(\tau_V+\tau_A)+
(1-{1 \over {N_C^2}})(\omega_V+\omega_A)
\end{equation}
\begin{equation}
{\bar f_B^2} M_B \epsilon_2 = \rho_2 = {1 \over {N_C}}
(\tau_V-\tau_A)-
{1 \over {2}}(1-{1 \over {N_C^2}})(\omega_V-\omega_A)
\end{equation}
 
In the case the $\Lambda_b$ baryon, the nonfactorisable
piece corresponds to
\cite{neubert97}
\begin{equation}
<(\bar b q)_{V-A}(\bar q b)_{V-A)} = -{1 \over {2N_C}}
\lambda_1-\lambda_2
\end{equation}

\section{Decay Rates and Lifetimes}
\label{sec:level1}
The decay rates of the b-flavoured hadrons are given by
\begin{equation}
\Gamma(H_b \rightarrow H_c l \bar \nu_l) =
{G_f^2 m_b^5 |V_{cb}|^2 \over {192 \pi^3}}
\left [(1+{\lambda_1+3\lambda_2 \over {2m_Q^2}})f(x)-
{6\lambda_2 \over {2m_Q^2}}f'(x)\right ]
\end{equation}
where, with $x = m_c^2/m_b^2$
\begin{equation}
f(x) = 1-8x+8x^3-x^4+12x^2ln x 
\end{equation}
\begin{equation}
f'(x) = (1-x)^4
\end{equation}
are the QCD phase space factors and $\lambda_1$ and $\lambda_2$  
correspond to the motion of the heavy quark inside the hadron and 
the chromomagnetic interaction respectively. These values are
taken to be -0.5 GeV for meson and -0.43 GeV for baryon and 0.12
GeV for meson. The chromomagnetic energy is zero for all
baryons except ${\Omega_Q}$.

Equation (44) is further supplemented by the FQO at 
the order (1/m$_Q^3$) in the HQE. At this order due to the light quarks
there are three processes: Pauli interference (PI),
Weak Annihilation (WI) and Weak Scattering (WS). The PI plays a
predictive role in the charged meson and the ${\Lambda_b}$ baryon.
The PI becomes constructive at the tree level whereas it becomes
destructive if radiative corrections are considered. The WS takes
place in the neutral meson as well as ${\Lambda_b}$.

\subsubsection{Lifetime ratio of B$^-$ and B$_d$}
Although the difference between the lifetimes of the charged and
the neutral B-mesons is almost a settled issue, we check the once
again using the expectation values of the colour-straight operators.
This difference is attributed to PI and WA. Neglecting the WA as
it is strongly CKM suppressed the result for  the PI is
\begin{equation}
\Delta\Gamma^f(B^-) = \Gamma_0 24 \pi^2 C_0{<O_V^q>_{B^-}-
<O_A^q>_{B^-} \over {m_b^3}}
\end{equation}
where C${_0}$ = ${c_+^2-c_-^2+{1 \over {N_c}}(c_+^2+c_-^2)}$ and 
${\Gamma_0}$ = ${G_f^2 m_b^5 |V_{cb}|^2 \over {192 \pi^3}}$.
The values for Wilson coefficients are: c${_+}$ = 0.84 and c${_-}$
= -1.42 with N${_c}$ = 3. For ${\Gamma_0}$, m${_b}$ = 4.8 GeV and
${|V_{cb}|}$ = 0.04 are used. Then the ratio is
\begin{equation}
{\tau(B^-)\over {\tau(B_d)}} = 1.00  
\end{equation}
This agrees well with the one obtained in terms B-meson decay
constants.

The decay rates due to spectator quark(s) processes are:
For B$^-$,
\begin{equation}
\Delta\Gamma^{nf}(B^-) = {G_f^2 m_b^2 \over {12\pi}}
|V_{cb}|^2(1-x)^2
[(2c_+^2-c_-^2)\phi_1+3(c_+^2+c_-^2)\rho_1]
\end{equation}
Hence the ratio is 1.03.

\subsubsection{Lifetime ratio of B$^-$ and B$_s$}
\label{sec:level3}
The lifetimes difference between the two neutral mesons
B$_s$ and B$_d$ is
due to $W$ exchange. The numerical result is 
\begin{equation}
{\tau(B_s) \over {\tau(B_d)}} = 1.00  
\end{equation}

Corresponding to the nonfactorisable part, we get the decay rate
\begin{equation}
\Delta\Gamma^{nf}(B_s) = {G_f^2 M_b^2 \over {12\pi}}
|V_{cb}|^2 (1-4x)^{1/2} 
[{(2c_+-c_-)^2 \over {3}}((1-x)\phi_{1}^s-(1+2x)\phi_{2}^s)+
{(c_++c_-)^2 \over {2}}((1-x)\rho_{1}^s-(1+2x)\rho_{2}^s)]
\end{equation}
Therefore the ratio becomes 1.02.

\subsubsection{Lifetime ratio of ${\Lambda_b}$ and B$^-$}
\label{sec:level3}
In the HQE, the difference in lifetimes between mesons and
baryons begins to appear at order 1/m$_Q^2$. Nevertheless it
is dominant at third power in 1/m$_Q$. At this order, the
FQO receives corrections due to WS and PI.
They are
\begin{eqnarray}
\Gamma_{WS}(\Lambda_b) = 92 \pi^2 \Gamma_0c_-^2
{<O_V^q>_{\Lambda_b}
\over {m_b^3}}\\
\Gamma_{PI}(\Lambda_b) = -48 \pi^2 \Gamma_0C_1
{<O_V^q>_{\Lambda_b}
\over {m_b^3}}
\end{eqnarray}
where $C_1 = -c_+(2c_--c_+)$. As mentioned earlier PI is
destructive for radiative corrections and enhances the
decay rate leading to smaller lifetime for ${\Lambda_b}$.
The effect of WS, on the other hand, is colour enhanced
and its consequence is smaller. Hence, 
\begin{equation}
{\tau(\Lambda_b) \over {\tau(B^d)}} = 0.79
\end{equation}

The deacy rate modified by the nonfactorisable piece is given by 
\begin{equation}
\Delta \Gamma (\Lambda_b) = {G_f^2 m_b^2 \over {16 \pi}}
\bar \lambda
[4(1-x)^2(c_-^2-c_+^2)-(1-x)^2(1+x)(c_--c_+)(5c_+-c_-)]
\end{equation}
where $\bar \lambda$ stands for the term in Eq. (43).
Correspondingly,
the ratio
is
\begin{equation}
{\tau(\Lambda_b) \over {\tau(B^-)}} = 0.84
\end{equation}
In mesonic cases, the nonfactorisable piece gives a little
bit higher  values. In particular, the ratio of
the lifetimes of the baryon and meson is significantly larger.  

\section{Conclusion}
In this paper, we have evaluated the FQO
for beauty hadrons. Though the spectator effects are suppressed
by powers of $({\Lambda_{QCD}/m_Q)^3}$, in the HQE for inclusive
decays, they cannot be neglected. We have expressed the four-quark
operators in terms of light quark scattering form factor which are
in turn related to the harmonic oscillator wave function. The use
of the wave function model is to replace the exponential and two
pole ansatz used in \cite{pirjol98}. Basically both are same.
The distinction arises only due to ${\beta}$, the oscillator
strength of the model. Interestingly this simple alternative
predicts the lifetimes ratio of ${\Lambda_b}$ and B closer
to the experimental value.

On the other hand, the nonfactorisable part does not have much
effect in the case of mesons. But it keeps still away the ratio
between B and ${\Lambda_b}$ away from the experimental value.
As far the B-mesons are concerned, the present study 
once again affirms the existing predictions.  In this case too,
there are omissions like SU(2) and SU(3) symmetry breaking.
They may play a role but too negligibly. 

Finally, we conclude that we have taken one, which is
dominant, of the sources of the preasymptotic effects and
shown that it predicts the lifetime of the ${\Lambda_b}$
close to the experimental figure. As we have not taken into
account all possible corrections to the four-quark
operators, the present prediction can be considered at
least indicative in order to look into the four-quark as
well as six-quark operators more seriously. However given
the basis provided in \cite{pirjol98}, the prediction has
to be believed. Of course, this prediction can be checked by
lattice studies. A refined analysis of b and c flavoured
baryon lifetimes will be published elsewhere.

\acknowledgements
The author is grateful to Prof. P. R. Subramanian for useful
discussion and constant encouragement. He thanks Dr R. Premanand
for clearing certain doubts in preparing this paper. The
University Grants Commission is thanked for its support through
Special Assistance Programme. He is thankful to F. De Fazio who
brought the Ref.\ \onlinecite{colan96} to his attention.
I thank the referee for his useful comments.

\end{document}